\documentclass[epj,twocolumn]{webofc}
\usepackage[varg]{txfonts}   
\newcommand{\be}{\begin{equation}}
\newcommand{\ee}{\end{equation}}
\newcommand{\bea}{\begin{eqnarray*}} 
\newcommand{\eea}{\end{eqnarray*}}
\newcommand{\beq}{\begin{eqnarray}}
\newcommand{\eeq}{\end{eqnarray}}
\newcommand{\Dlr}{\buildrel \leftrightarrow \over D\raise-1pt\hbox{}}

\wocname{Meson-Nucleon Physics and the Structure of the Nucleon}
\woctitle{The 13${\rm th}$ International Conference on  Meson-Nucleon Physics and the Structure of the Nucleon}
\begin{document}
\title{Nucleon structure from lattice QCD - recent achievements and perspectives}
%
%

\author{Constantia Alexandrou\inst{1,2}\fnsep\thanks{\email{alexand@ucy.ac.cy}} 
}

\institute{Department of Physics, University of Cyprus, PO Box 20537, 1678 Nicosia, Cyprus
\and
          Computation-based Science and Technology Research Center, The Cyprus Institute, P.O. Box 27456, 1645 Nicosia, Cyprus}

\abstract{
We present recent developments in lattice QCD simulations as applied in the study of hadron structure. We discuss the challenges and perspectives in the evaluation
of benchmark quantities such as the nucleon axial charge and the isovector parton momentum fraction, as well as, in the computation of the nucleon $\sigma$-terms, which involve
the calculation of disconnected quark loop contributions.
}
\maketitle
\section{Introduction}
\label{intro}
There has been spectacular progress in lattice QCD simulations during the past few years, resulting from improvements in algorithms and faster computers.
The latest important development is the simulation of the full theory with
quark masses tuned to their physical values. Gauge field configurations
are now available at near  physical pion mass for Wilson-type, staggered and domain wall fermions.  This enables us to compute physical quantities directly at
the physical point avoiding systematic errors due to the  chiral extrapolation, which for baryons are particularly difficult to reliably estimate.

\begin{figure}
\includegraphics[width=\linewidth]{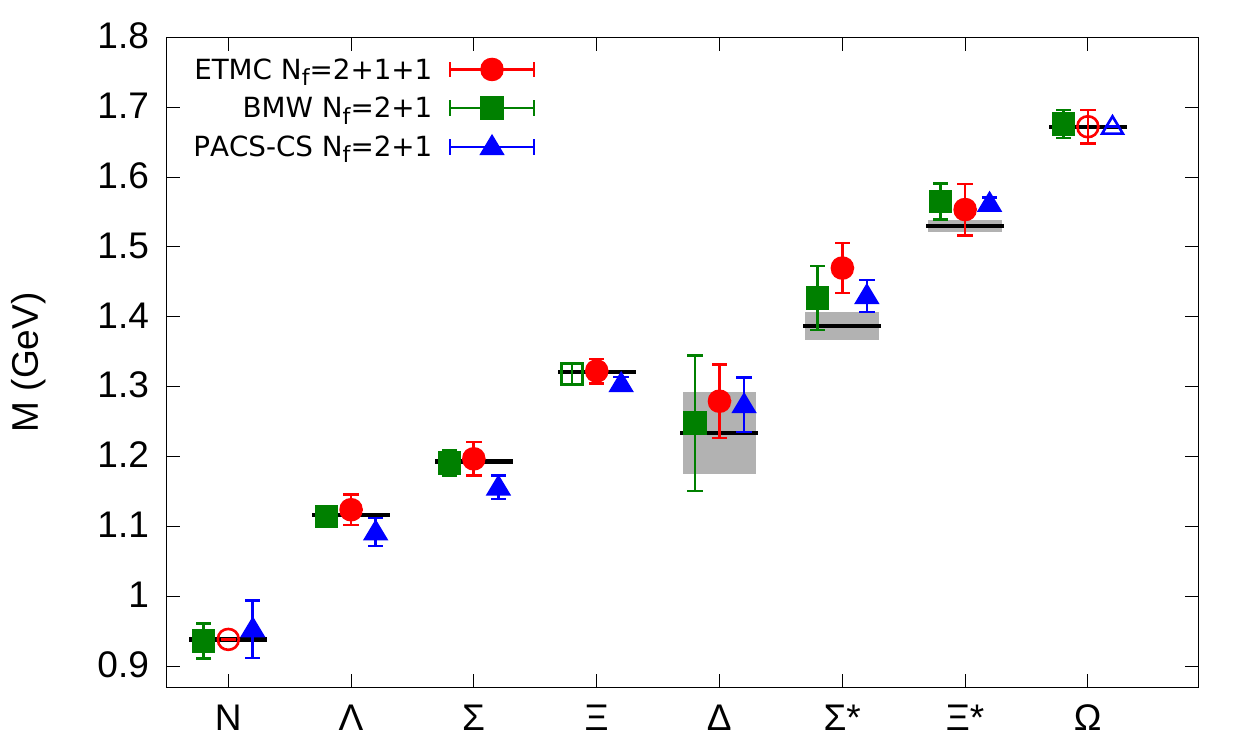}\\
 \includegraphics[width=\linewidth]{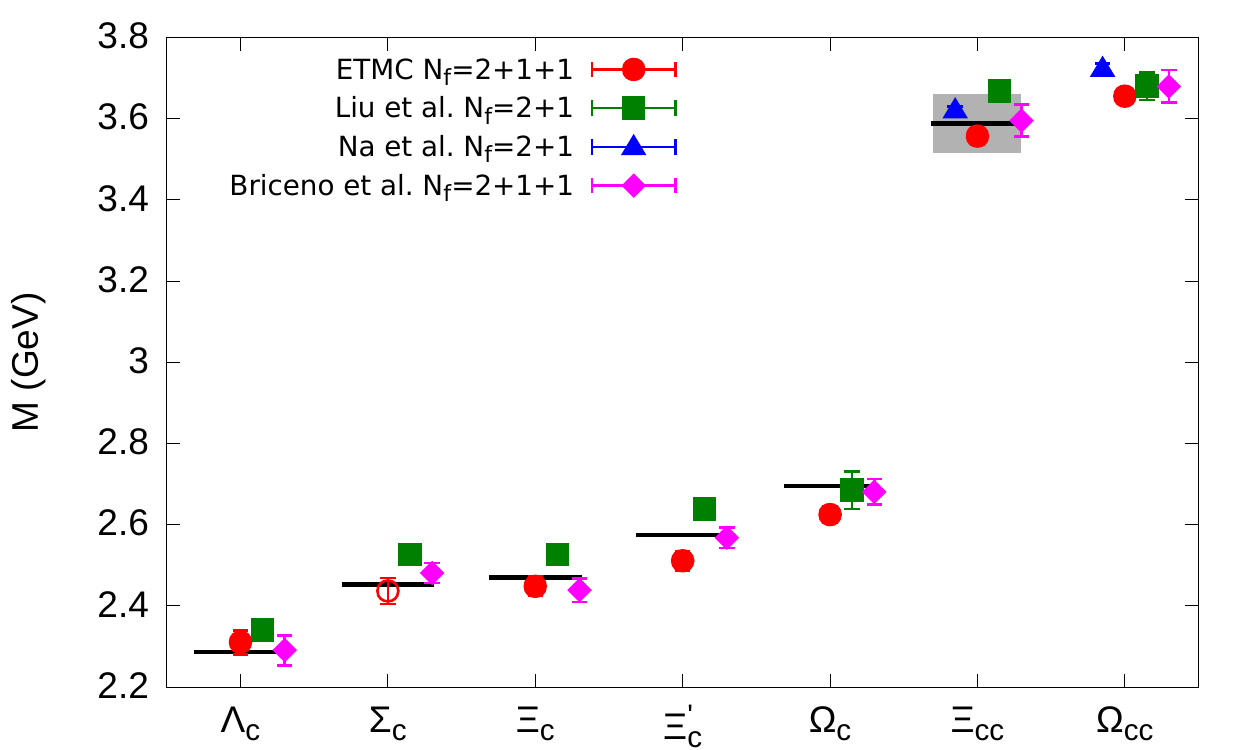}\\
 \includegraphics[width=\linewidth]{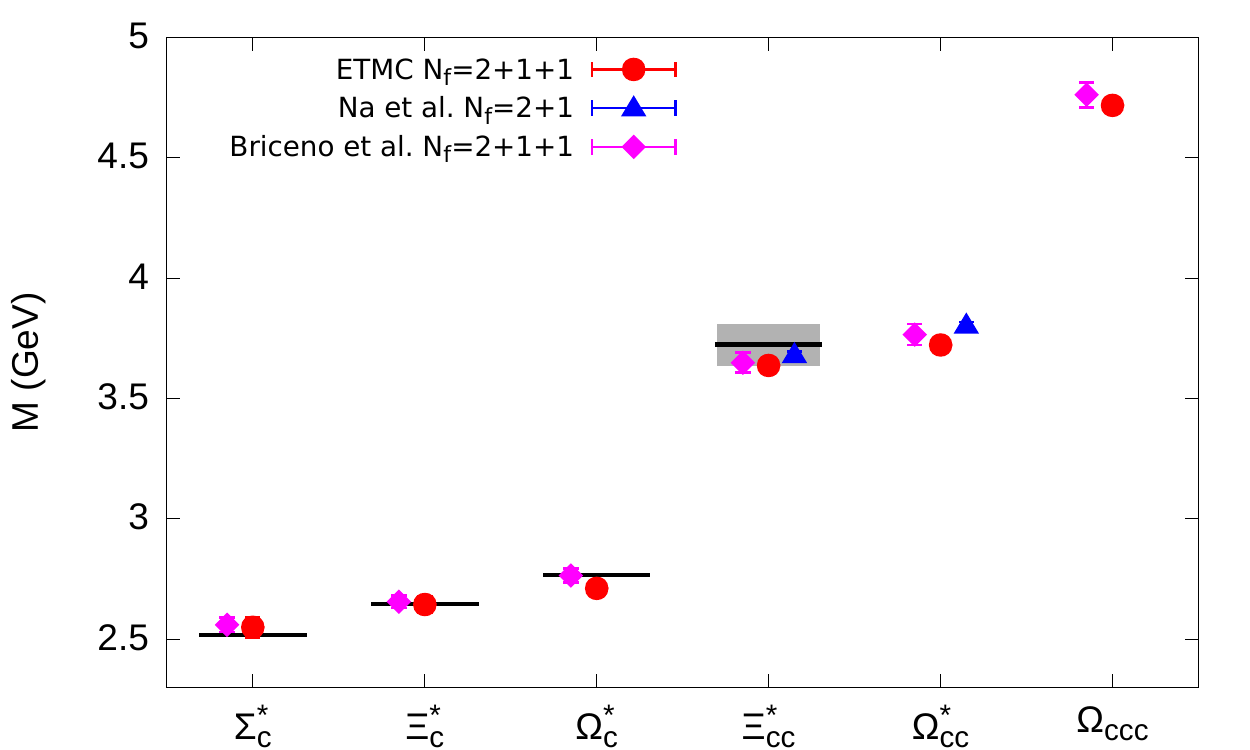}
\label{fig:mass}
\caption{Upper: The mass of octet and decuplet baryons. Results are by the BMW~\cite{Durr:2008zz} and the ETM collaborations, extrapolated to the continuum limit and by the PACS-CS at one lattice spacing~\cite{Aoki:2008sm}. Open symbols show input quantities used to determine the lattice spacing and strange quark mass; middle: The mass of 1/2-spin charmed baryons using twisted mass fermions and staggered sea quarks~\cite{Briceno:2012wt,Liu:2009jc,Na:2008hz}; Lower: The mass of 3/2 spin charmed baryons using twisted mass fermions and staggered sea quarks~\cite{Briceno:2012wt,Liu:2009jc}. The results by the ETM collaboration are preliminary since only statistical errors are shown.}
\vspace*{-0.5cm}  
\end{figure}

Reproducing the low-lying hadron spectrum has been a milestone for lattice QCD.
 In Fig.~\ref{fig:mass}
we show the pioneering  results for the masses of the octet and decuplet baryons 
produced by the BMW collaboration using
 $N_f=2+1$ clover fermions~\cite{Durr:2008zz}, as well as, results by the ETM collaboration using $N_f=2+1+1$ twisted mass fermions (TMF)~\cite{CA:future}. Both collaborations employed simulations with pion masses ranging from  about 200~MeV to 500~MeV and extrapolated to the continuum limit using three lattice spacings. In addition, results from the PACS-CS collaboration using $N_f=2+1$ clover fermions at one lattice spacing of $a=0.0907(13)$~fm~\cite{Aoki:2008sm} are included. As can be seen, lattice QCD results extrapolated to the continuum limit, taking into account systematic errors as performed by the BMW collaboration are in agreement with the experimental values.  In Fig.~\ref{fig:mass} we also show {\it preliminary} results using $N_f=2+1+1$ twisted mass fermions  for the spin-1/2 and -3/2 charmed baryons extrapolated to the continuum limit. 
The  mass of the $\Sigma_c$ baryon is used to fix the mass of the charm quark obtaining a value that is in agreement with the one extracted from the D-meson mass. Besides the statistical errors shown,  systematic errors arising from the tuning of the charm quark mass and the chiral extrapolation are currently being assessed. Results on charmed baryons  obtained  using staggered gauge field configurations are also shown in Fig.~\ref{fig:mass}. In Ref.~\cite{Briceno:2012wt}  $N_f=2+1+1$ staggered sea quarks with clover light and strange valence quarks  and a relativistic action for the charm quark are employed and the results are  extrapolated to the continuum limit. In Refs.~\cite{Na:2008hz,Liu:2009jc}  $N_f=2+1$ staggered sea quarks are used with  staggered light and strange~\cite{Na:2008hz} or domain wall~\cite{Liu:2009jc}   valence quarks with a relativistic action for the charm quark.
These results on the hadron masses demonstrate  that  the known spectrum including the mass of charmed baryons can be reproduced within lattice QCD thus enabling lattice QCD to provide predictions for the masses of  those that have not been measured.

\section{Challenges and  perspectives}

In what follows we survey results on benchmark  observables, which are still a challenge for lattice QCD and discuss on-going efforts to evaluate systematic
errors that may lead to their understanding.  
\subsection{Masses of excited states}
\begin{figure}
\includegraphics[width=\linewidth]{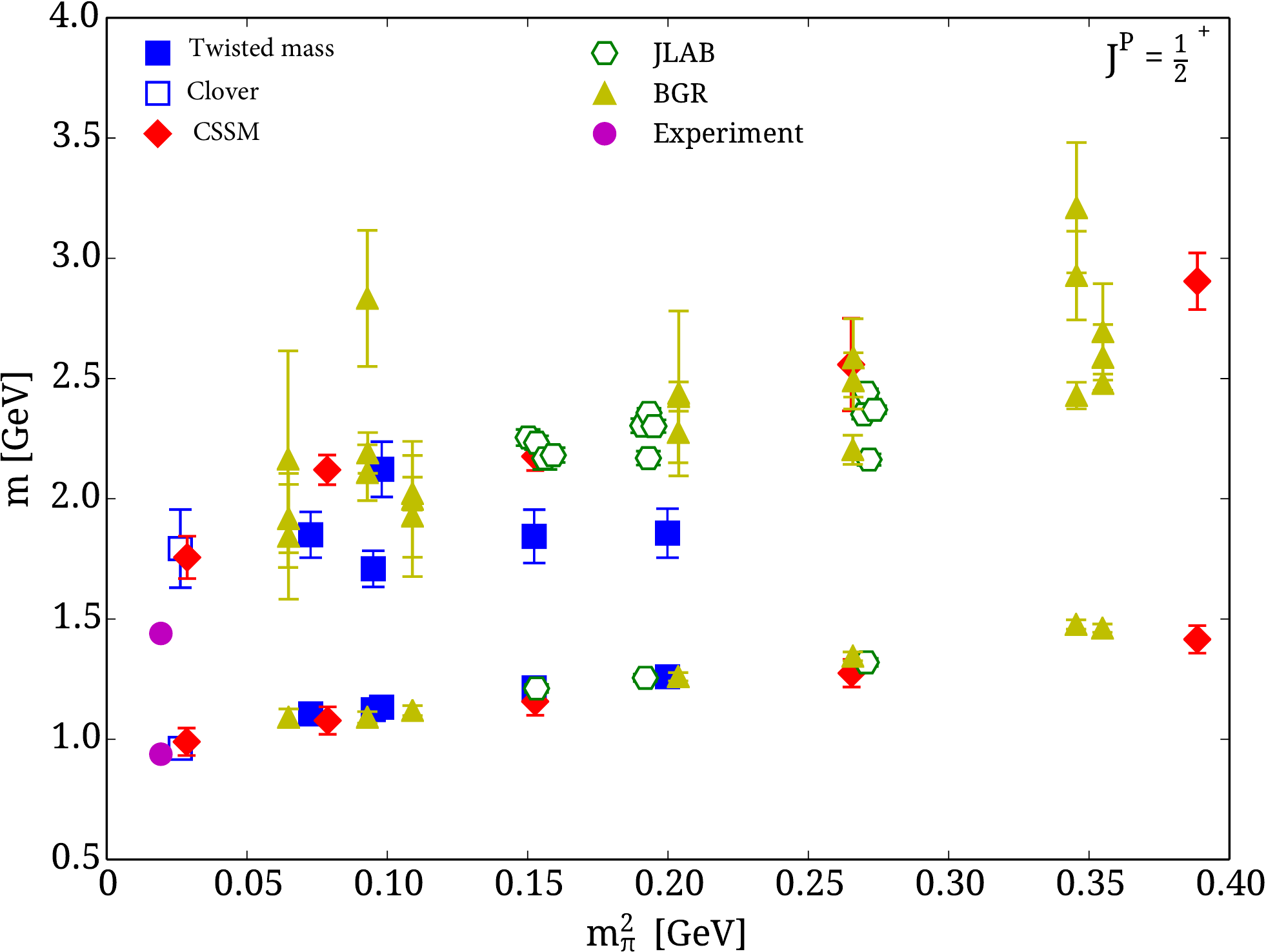}\\
\includegraphics[width=\linewidth]{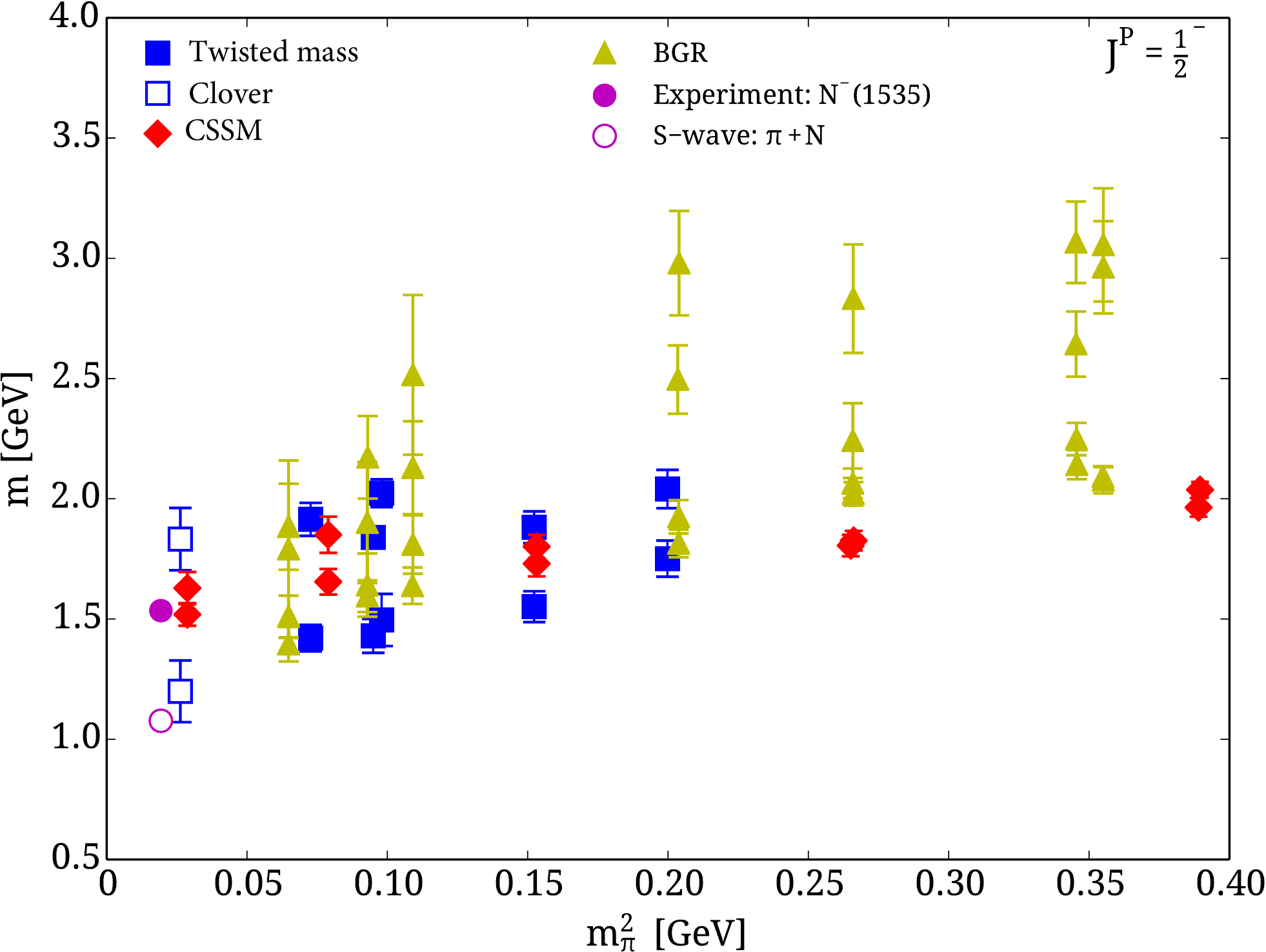}
\caption{The two lowest nucleon states in the positive (upper) and negative (lower) parity channels. Results are
shown using twisted and clover fermions from Ref.~\cite{Alexandrou:2013fsu}, using clover fermions from Refs.~\cite{Mahbub:2012zz,Mahbub:2013ala,Edwards:2011jj} and using a chirally improved Dirac operator from Ref.~\cite{Engel:2013ig}. In the negative parity channel the lowest $\pi-N$ scattering state is identified.}
\label{fig:excited}
\end{figure}

While Euclidean correlation functions are very well suited for studies of the lowest
hadron state, extracting  excited states is harder since they are exponentially suppressed as compared to the ground state.  A variational approach, where we enlarge the basis of interpolating fields is the standard approach employed in order to obtain excited states.
We thus consider the correlation matrix 
\be 
G_{jk}(\vec q, t_s)=\sum_{{\vec x}_s} \, e^{-i\vec {x}_s \cdot \vec q}\, 
     \langle{J_j(\vec {x}_s,t_s)J_k^\dagger(0)} \rangle\>, j,k=1,\ldots N 
\ee
and  solve the generalized eigenvalue problem:

\be 
G(t)v_n(t;t_0)=\lambda_n(t;t_0) G(t_0)v_n(t;t_0),\,\, \lambda_n(t;t_0)=e^{-E_n(t-t_0)}
\label{GEVP}
\ee
 that yields the N lowest eigenstates~\cite{Luscher:1990ck}.
Large effort has been devoted to construct  appropriate bases using lattice symmetries~\cite{Morningstar:2013bda} that also includes multi-hadron states. Besides using a suitable variational basis one needs  to consider disconnected diagrams as well as develop methods to
deal with resonances since most  excited states are unstable at the physical pion mass. In this presentation we limit ourselves to examining the first two excited states of the nucleon
and in particular the Roper~\cite{Alexandrou:2013fsu,Alexandrou:2013cda}.
 In the positive
parity channel  a linear
combination of interpolating fields corresponding to a small and large
root mean square radius (rms) produces a wavefunction with a node
having potentially a larger overlap with the Roper state. We indeed
observe a lowering in the energy of the first excited state when
including an interpolating field with a large rms radius. In Fig.~\ref{fig:excited} we show results on the two lowest states in the positive and negative parity
channels. The energy of the $\pi-N$ scattering state is clearly shown in the negative parity channel. However, the results on the Roper and on $S_{11}(1535{\rm MeV})$ generally have larger errors and a systematic study is still required to reach a 
definite conclusion. 

  \subsection{ Nucleon  form factors}

\begin{figure}
 \centerline{\includegraphics[width=0.7\linewidth]{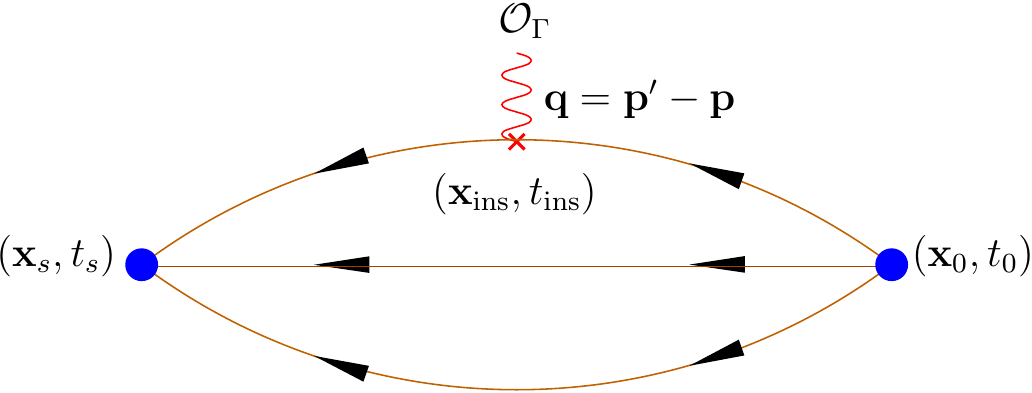}}\vspace*{0.3cm}
\centerline{\includegraphics[width=0.7\linewidth]{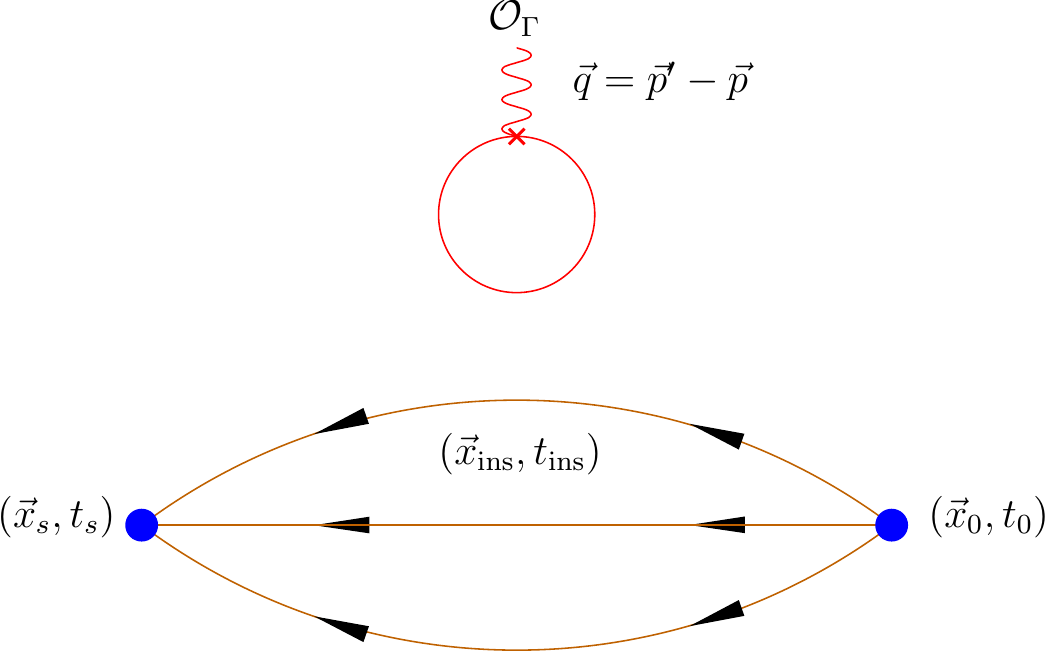}}
\caption{Connected (upper)  and disconnected (lower) contributions to the three-point function.}
\label{fig:conn and disconn}
\end{figure}
\begin{figure}
\centerline{\includegraphics[width=0.9\linewidth]{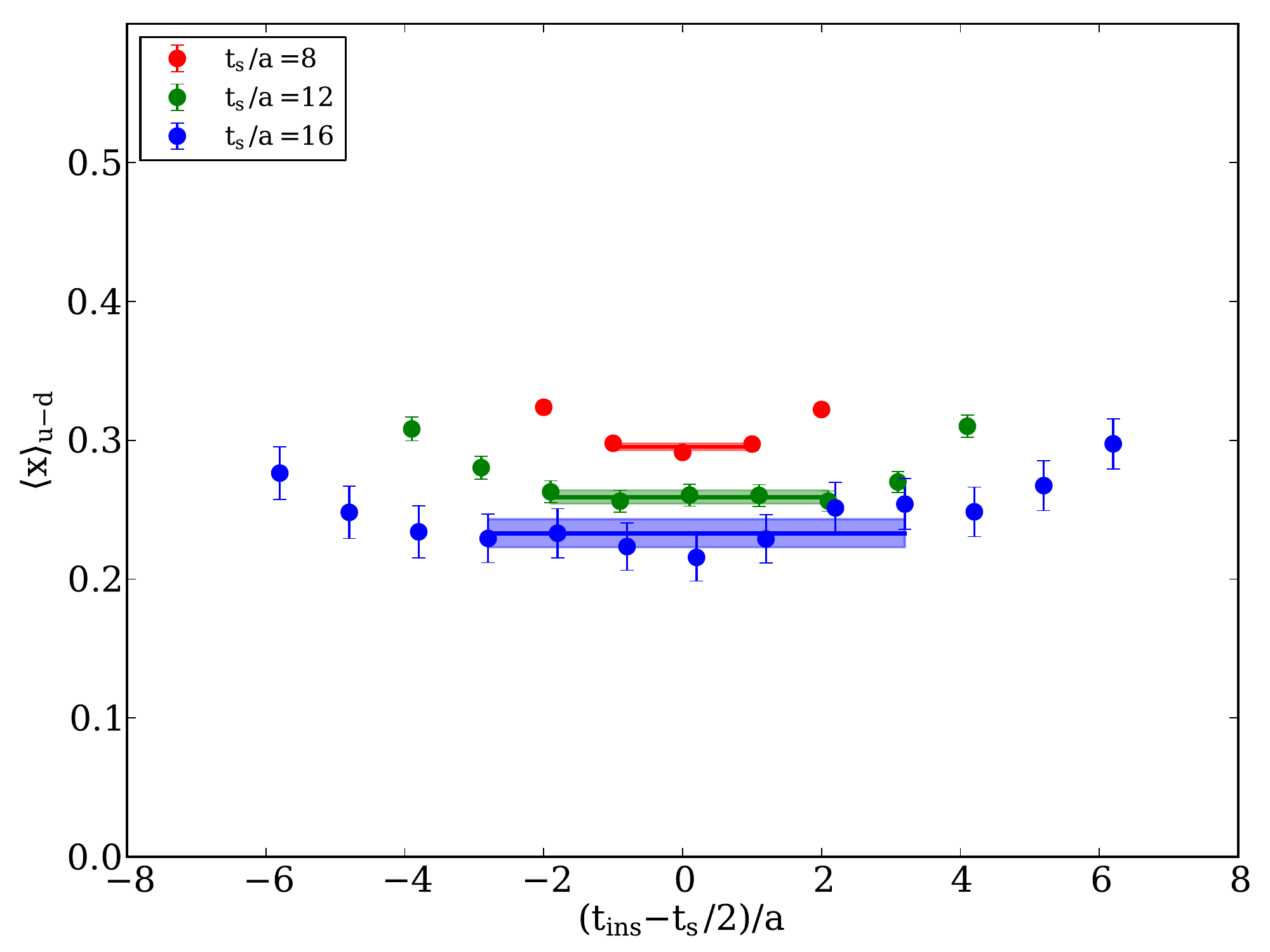}}
\label{fig:plateaus}
\caption{The ratio from which the isovector quark momentum fraction is extracted versus $t_{\rm ins}$ for sink-source time separations $t_s-t_0=8a,\, 10a,\, 12a$.}
\end{figure}
\begin{figure}
\includegraphics[width=\linewidth]{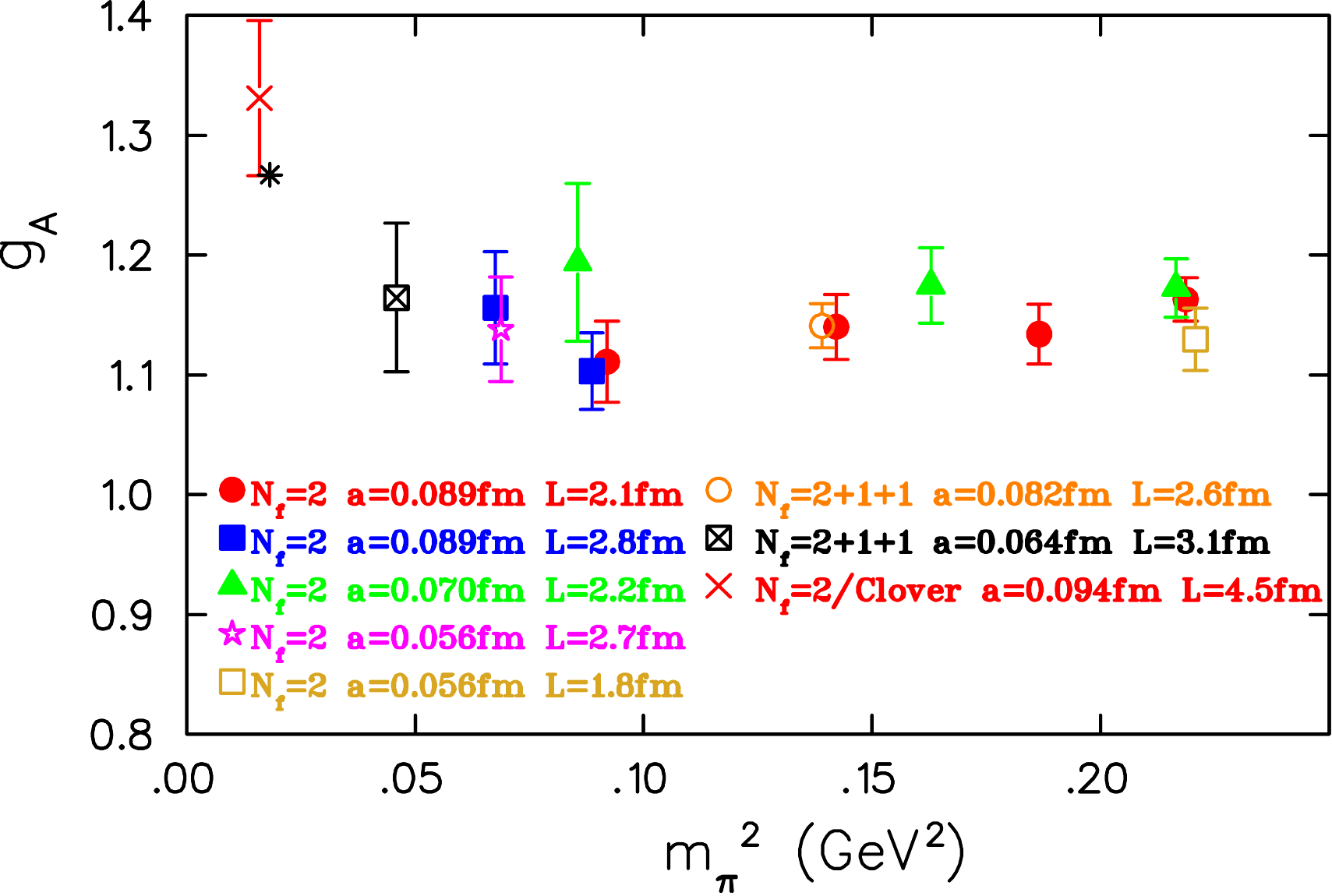}\\
\includegraphics[width=\linewidth]{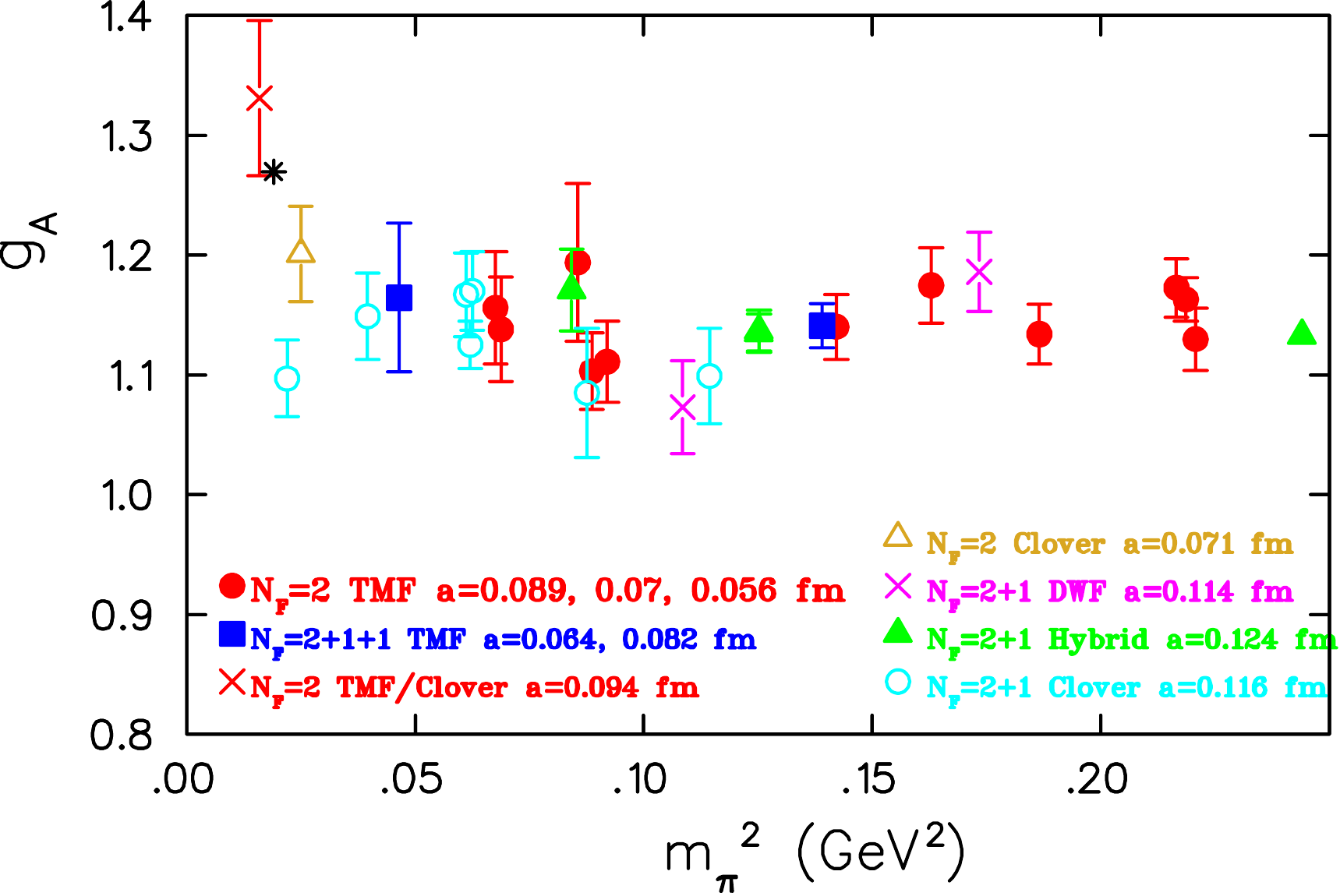}
\caption{Upper: $N_f=2$ and $N_f=2+1+1$ results with twisted mass fermions for three lattice spacings and different volumes~\cite{Alexandrou:2013jsa}. Lower: Comparison of lattice results extracted from the plateau value of $R(t_s,t_{\rm ins},t_0)$ using TMF,  $N_f=2+1$ clover fermions~\cite{Green:2012ud,Owen:2012ts}, $N_f=2$ clover fermions~\cite{Horsley:2013ayv,Capitani:2012gj} and a mixed action approach~\cite{Bhattacharya:2013ehc}. }
\label{fig:gA}
\end{figure}

\begin{figure}
 \includegraphics[width=0.93\linewidth]{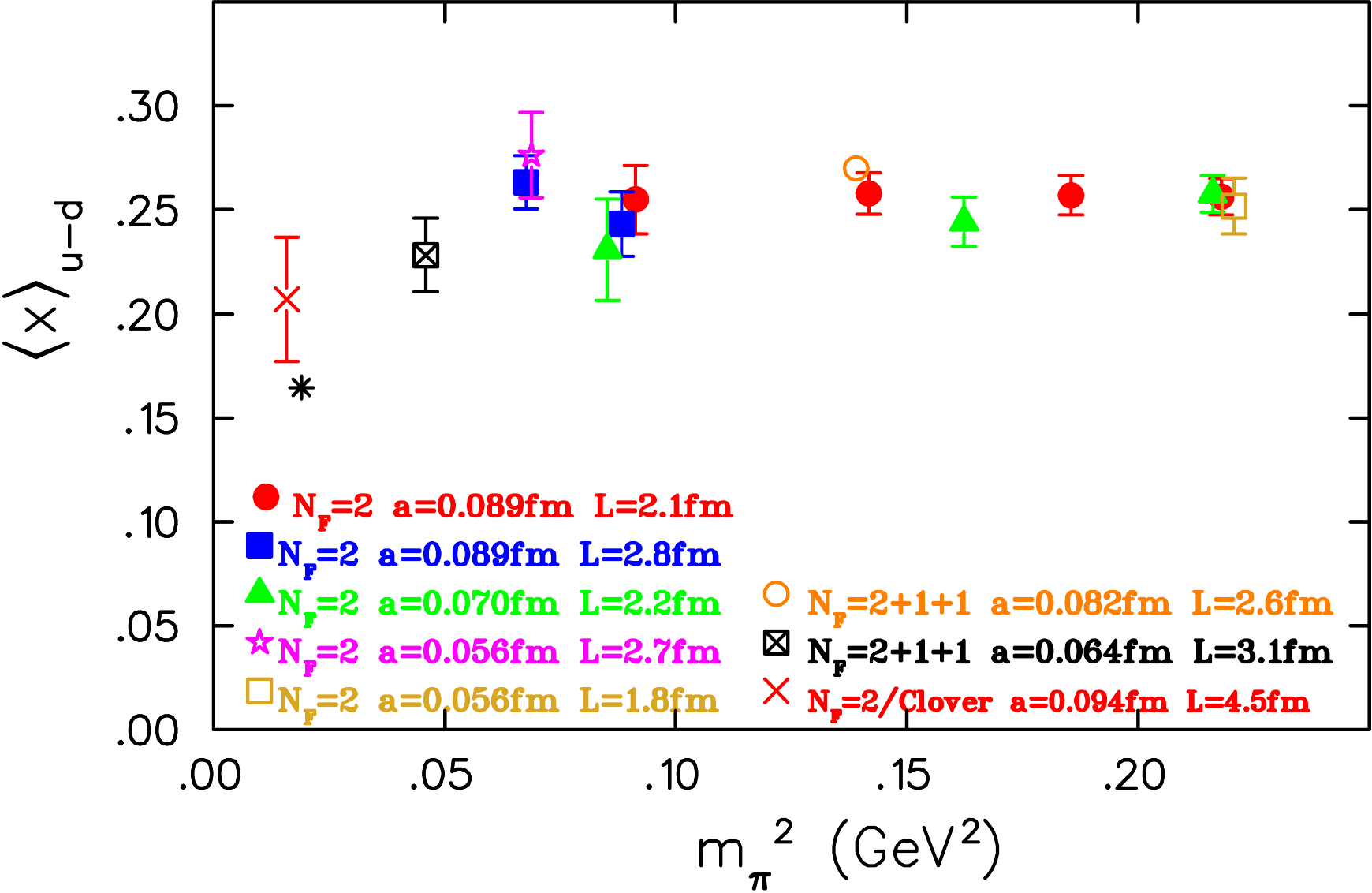}\\
 \includegraphics[width=0.93\linewidth]{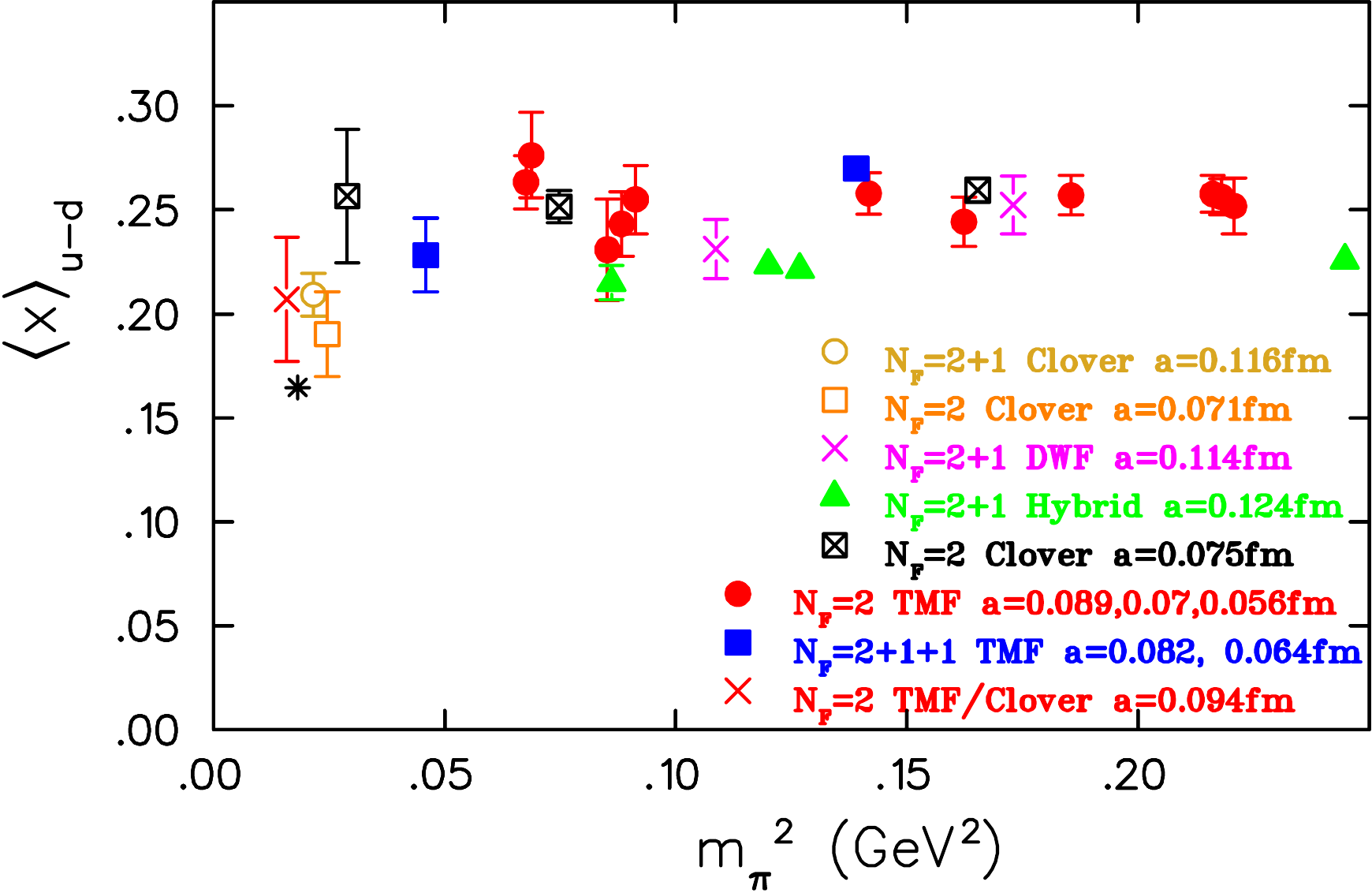}
\caption{The isovector momentum fraction $\langle x \rangle _{u-d}$ versus $m_\pi^2$. Results using twisted mass fermions (upper)~\cite{Alexandrou:2013jsa} are shown together with a comparison (lower) with results using $N_f=2$ clover fermions~\cite{Bali:2012av,Pleiter:2011gw}, $N_f=2+1$ clover fermions~\cite{Green:2012ud},  $N_f=2+1$ domain wall fermions~\cite{Aoki:2010xg} and within a hybrid approach of $N_f=2+1$ staggered sea and domain wall valence~\cite{Bratt:2010jn}. The experimental value is taken from Ref.~\cite{Alekhin:2012ig}.}
\label{fig:x}
\end{figure}

\begin{figure}
 \includegraphics[width=0.9\linewidth]{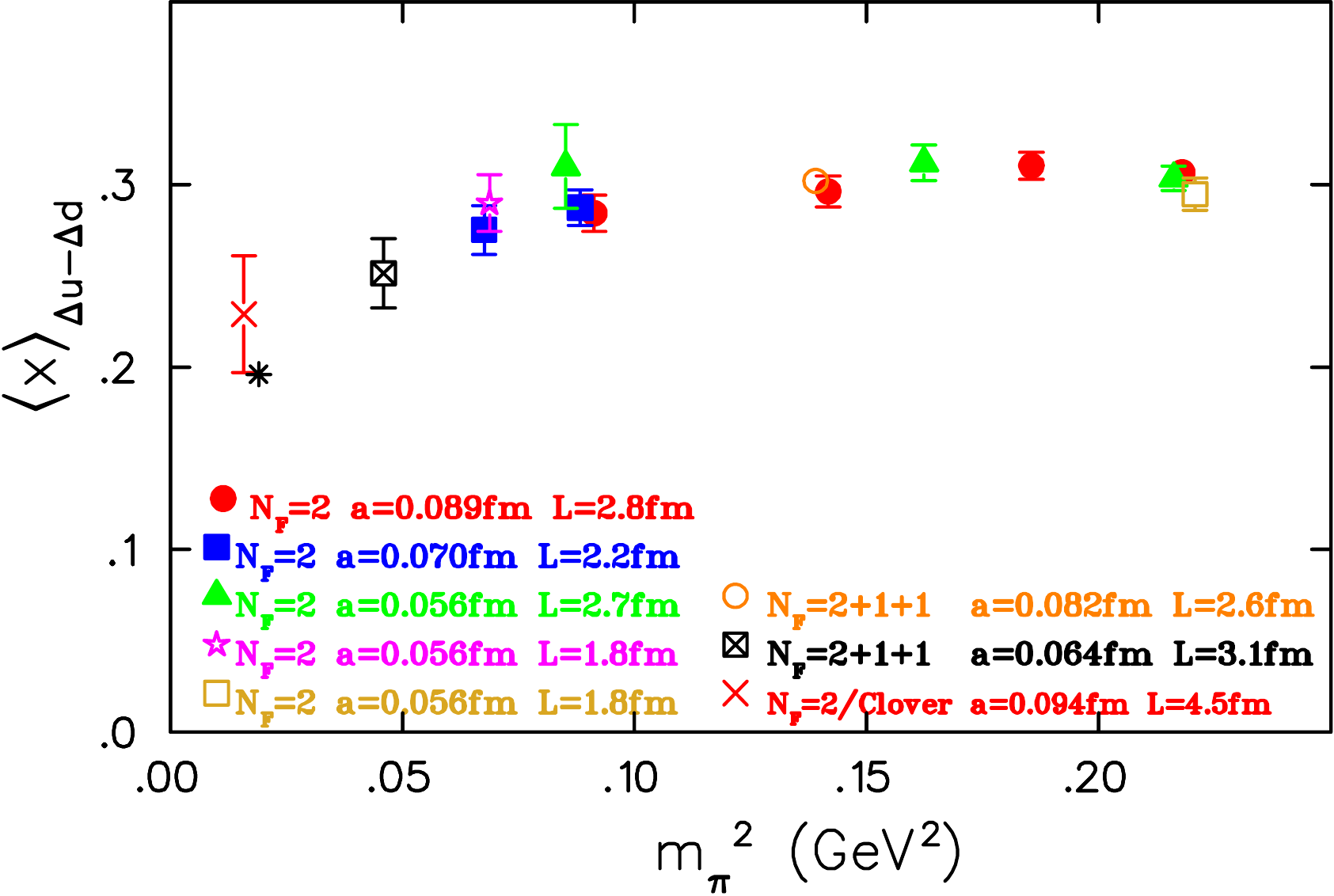}\\
 \includegraphics[width=0.9\linewidth]{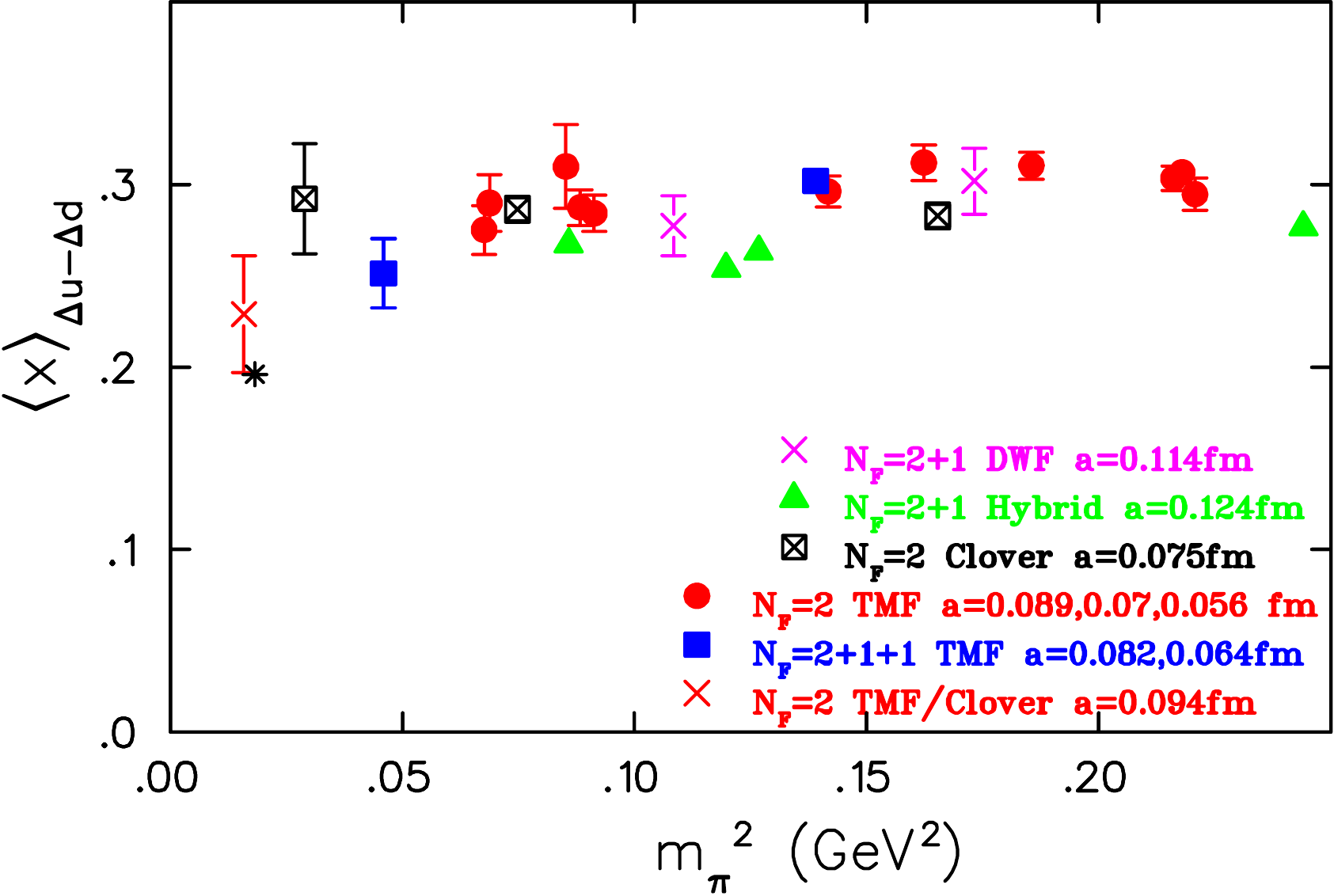}
\caption{The isovector helicity fraction $\langle x \rangle _{\Delta u-\Delta d}$ versus $m_\pi^2$. Results using twisted mass fermions (upper)~\cite{Alexandrou:2013jsa} are shown together with a comparison (lower) with results using $N_f=2$ clover fermions~\cite{Pleiter:2011gw},  $N_f=2+1$ domain wall fermions~\cite{Aoki:2010xg} and within a hybrid approach of $N_f=2+1$ staggered sea and domain wall valence~\cite{Bratt:2010jn}. The experimental value is taken from Ref.~\cite{Blumlein:2010rn}.}
\label{fig:Dx}
\end{figure}

\begin{figure}
{\includegraphics[width=0.9\linewidth]{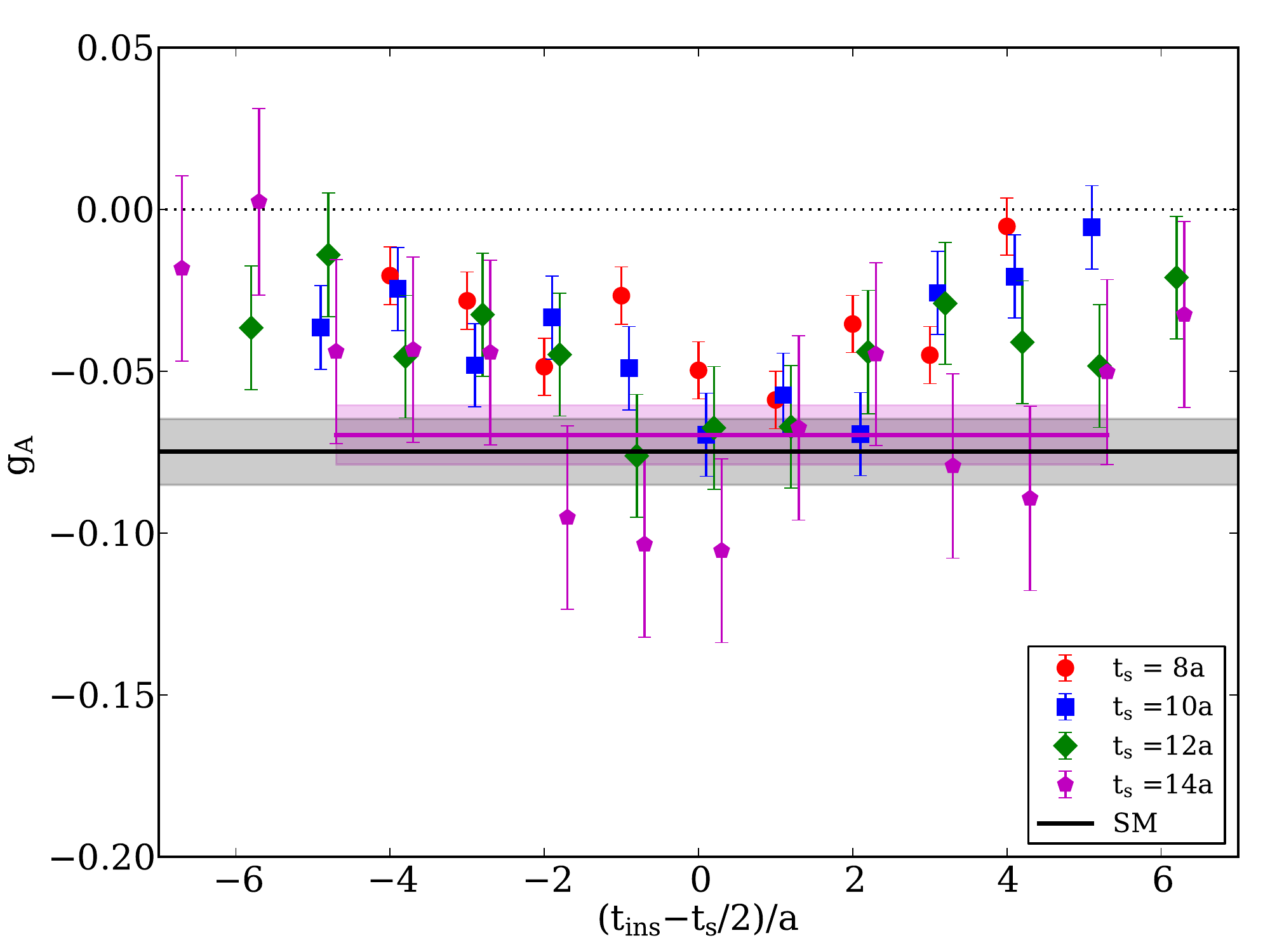}}\\
{\includegraphics[width=0.9\linewidth]{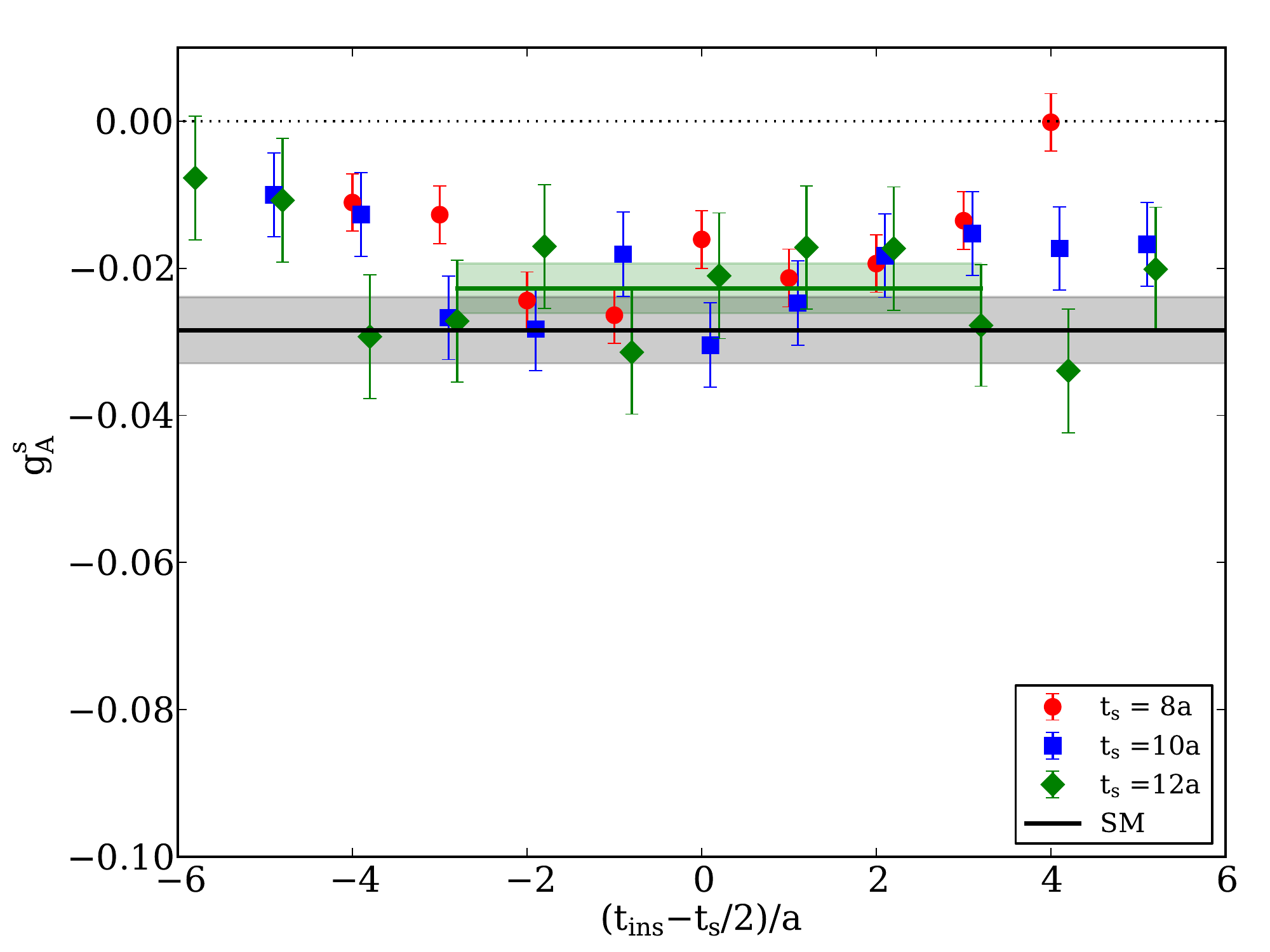}}
\caption{Upper: Disconnected contribution to the ratio from where the  isoscalar $g_A^{u+d}$ is extracted; Lower: The ratio from which the strange quark contribution $g_A^s$ is extracted. }
\vspace*{-0.5cm}
\label{fig:gA disc}
\end{figure}

\begin{figure}
{\includegraphics[width=\linewidth]{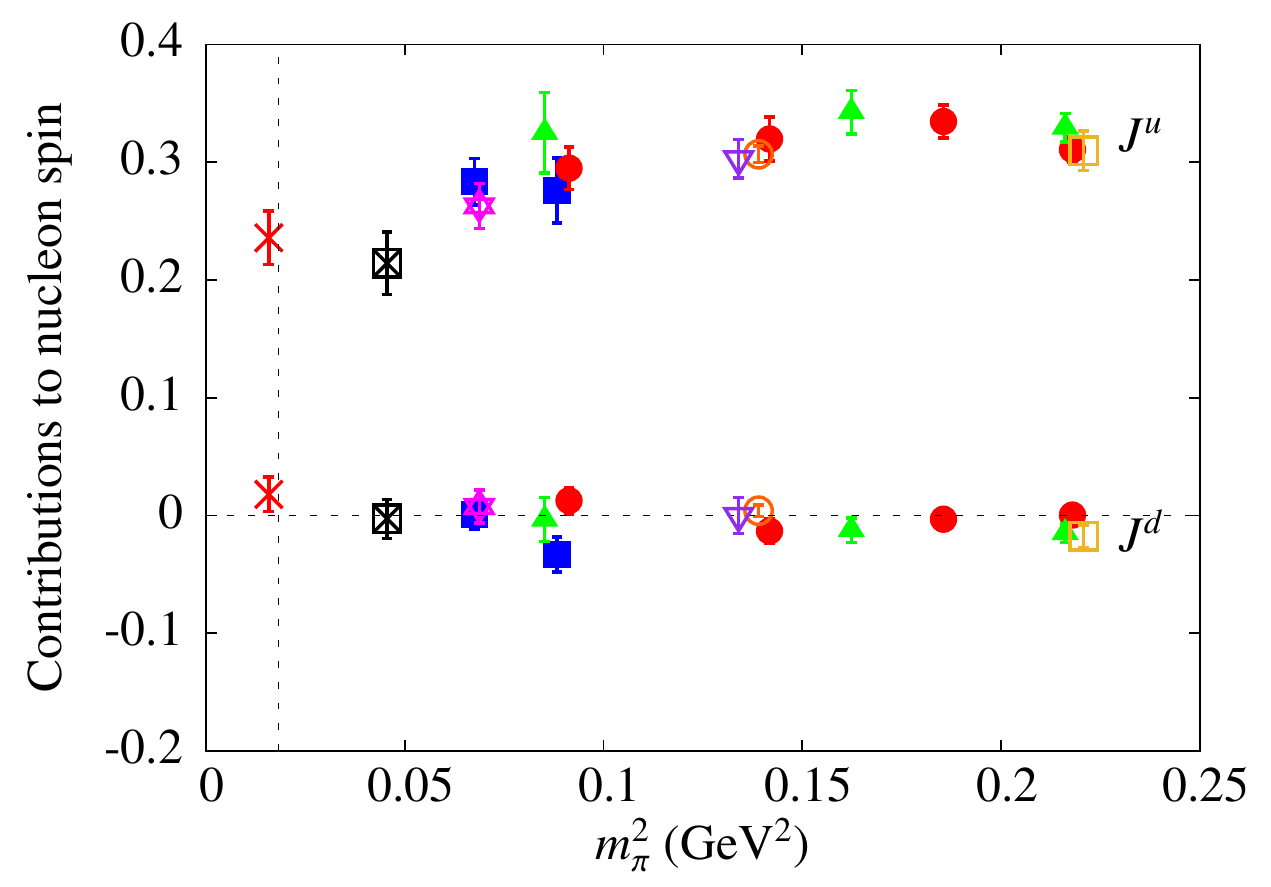}}\\
{\includegraphics[width=\linewidth]{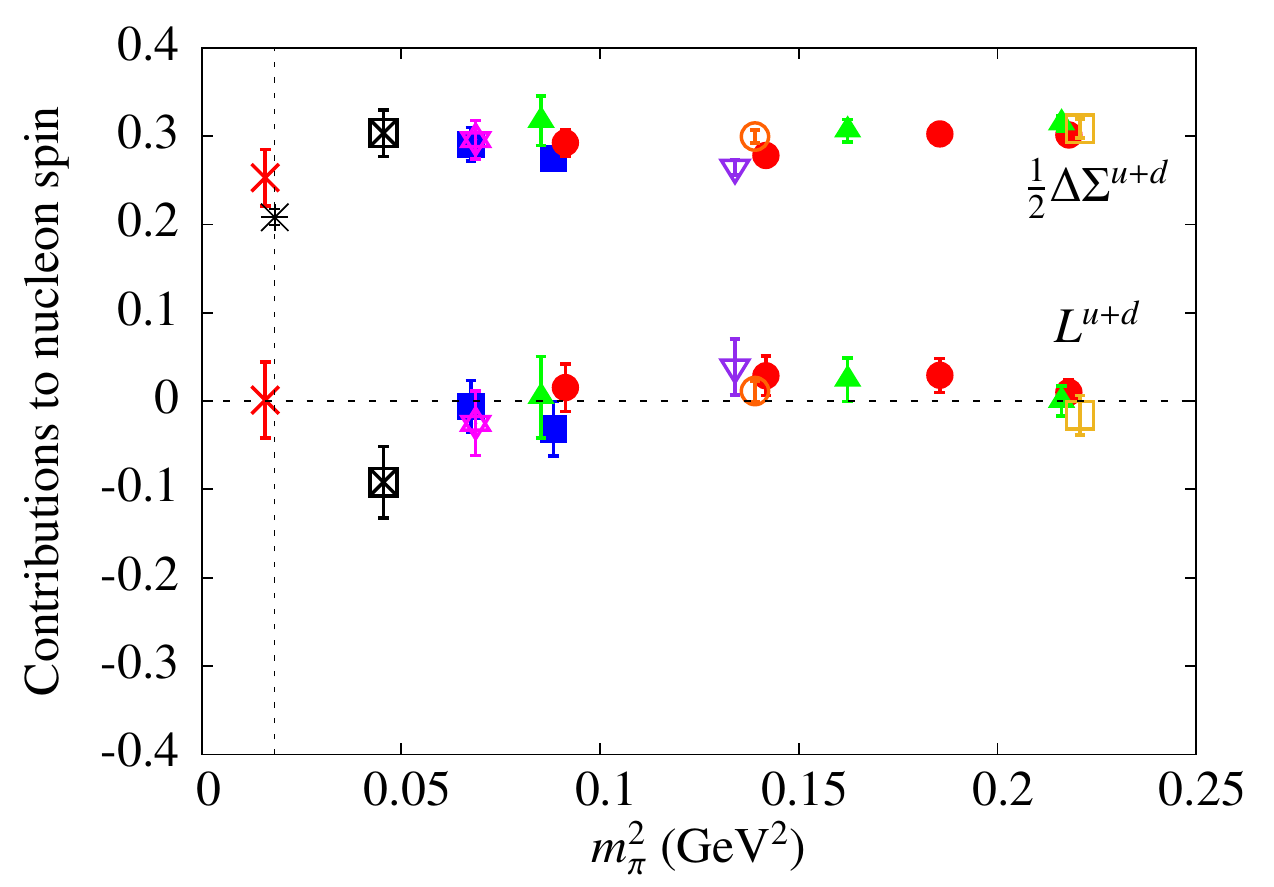}}
\caption{ $J^u$ and $J^d$ (upper) and $\Delta \Sigma^{u+d}$ and $L^{u+d}$ (lower) as a function of the pion mass. At $m_\pi=373$~MeV
we show the results obtained when we include the disconnected contributions (open inverted purple) triangle  using 150,000 measurement as compared to 1200 for the connected part (open red circle).} 
\label{fig:quark spin}
\end{figure}

\begin{figure}
\includegraphics[width=\linewidth]{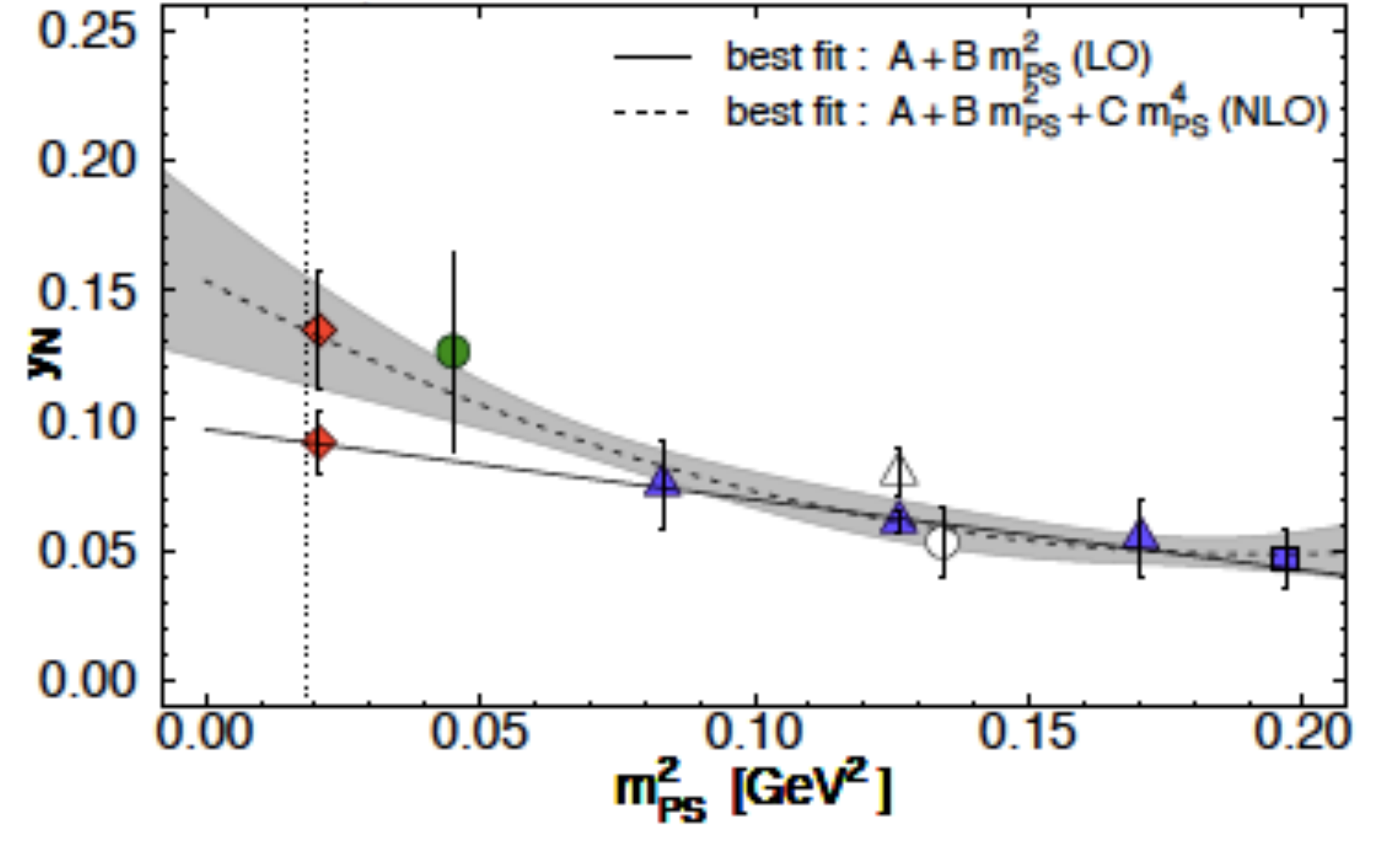}
\caption{The $y_N$-parameter versus $m_\pi^2$ together with extrapolations to the physical pion mass.}
\label{fig:sigma}\vspace*{-0.5cm}
\end{figure}

 The evaluation of the connected contribution to the three-point function 
$$ G^{\mu\nu}({\Gamma},\vec q,t_s, t_{\rm ins}) =\sum_{\vec x_s, {\vec x}_{\rm ins}} \, e^{i{\vec x}_{\rm ins} \cdot \vec q}\, 
     {\Gamma}\, \langle J(\vec x_s,t_s) {\cal O}^{\mu\nu}({\vec x}_{\rm ins},t_{\rm ins}) \overline{J}(\vec{x}_0, t_0) \rangle 
$$
shown schematically in Fig.~\ref{fig:conn and disconn}, is customarily carried out  via sequential inversions through the sink.
An appropriate ratio of the three-point function with nucleon two-point functions is constructed such that the exponential time decay due to the Euclidean time evolution and unknown overlaps cancel. This ratio behaves as
\small
 \be
    R(t_s,t_{\rm ins},t_0) \xrightarrow[(t_s-t_{\rm ins})\Delta \gg 1]{(t_{\rm ins}-t_0)\Delta \gg 1} \mathcal{M}[1
      + \dots e^{-\Delta({\bf p})(t_{\rm ins}-t_0)} + \dots e^{-\Delta({\bf
          p'})(t_s-t_{\rm ins})}]
  \ee
\normalsize
where $ \mathcal{M}$ the desired matrix element determined from the time-independent value  of $R(t_s,t_{\rm ins},t_0)$ (plateau value) as illustrated in Fig.~\ref{fig:plateaus}, $t_s,\>t_{\rm ins}$ and $t_0$ are the
  sink, operator insertion and source time-slices
and $\Delta({\bf p})$ the
  energy gap of the lowest state with the first excited state.
 Lattice results are connected  to the measured quantities through a multiplicative renormalization ${\cal O}_{\overline{\rm MS}}(\mu)=Z(\mu,a){\cal O}_{\rm latt}(a)$
with  $Z(\mu,a)$ computed  non-perturbatively.

\subsubsection{Axial charge $g_A$}

The nucleon matrix element of the axial-vector current $ A^3_\mu=\bar{\psi}\gamma_\mu\gamma_5 \frac{\tau^3}{2}\psi(x)$
can be written in terms of two  form factors as
\be
  \frac{1}{2}\bar{u}_N(\vec{p^\prime})\left[ \gamma_\mu\gamma_5 {G_A(q^2)} + \frac{q^\mu\gamma_5}{2 m} {G_p(q^2)} \right]u_N(\vec{p})|_{q^2=0}\ee
 yielding the nucleon axial charge  $G_A(0)\equiv g_A$.
$g_A$ is well-measured  and has no quark loop contributions and as such it constitutes a benchmark quantity for hadron structure calculations.
In Fig.~\ref{fig:gA} we show results using twisted mass fermions and provide 
a comparison of various lattice results extracted from determining the plateau value ${\cal M}$. 
We note that a number of collaborations are now producing results at or near physical pion mass. We expect that  a dedicated study  with high statistics, larger volumes and simulations at 3 lattice spacings will be  needed  in order to finalize these results.
Such studies are underway and lattice QCD is poised to resolve the discrepancy
on the value of $g_A$.

\subsubsection{Momentum fraction}
Another important quantity measured in deep inelastic scattering is the quark momentum fraction  
$\langle x \rangle_q=\int_0^1 dx x\left[q(x)+\bar{q}(x)\right]$ as well as the helicity moment 
$\langle x\rangle_{\Delta q}=\int_0^1 dx x\left[\Delta q(x)-\Delta \bar{q}(x)\right]
$
where $ q(x)=q(x)_\downarrow+q(x)_\uparrow$ and
 $\Delta q(x)=q(x)_\downarrow-q(x)_\uparrow$.
In lattice QCD these can be extracted  
by computing the nucleon matrix elements of $ {\cal O}_q^{\mu_1 \mu_2}  = \bar \psi  \gamma^{\{\mu_1}i\Dlr  ^{\mu_{2}\}} \psi$ and $ {\cal O}_{\Delta q}^{\mu_1 \mu_2}  = \bar \psi  \gamma^{\{\mu_1}\gamma_5i\Dlr ^{\mu_{2}\}} \psi$.
The results on the isovector $\langle x \rangle_{u-d}$ and $\langle x \rangle_{\Delta u-\Delta d}$ in the $\overline{MS}$ scheme at  $\mu= 2$~GeV are summarized in 
Figs.~\ref{fig:x} and \ref{fig:Dx}.
As can be seen, results obtained at or near the physical pion mass are converging to the experimental value and, like for the case of $g_A$, further studies are
expected to resolve the remaining discrepancies.

\section{Disconnected quark loop contributions}

The disconnected quark loop contributions to hadron matrix elements shown schematically in Fig.~\ref{fig:conn and disconn} are notoriously difficult to compute for two reasons: i) they are given by
 $L(x_{\rm ins})=Tr\left [\Gamma G(x_{\rm ins};x_{\rm ins})\right]\>$,
which  involves the quark propagator from all $\vec{x}_{\rm ins}$ (i.e. $L^3$ more inversions as compared to hadron masses)  and ii) they are prone to 
large gauge noise needing  large statistics.
To  compute such contributions to sufficient accuracy   special techniques are utilized that combine 
i) usage of stochastic noise  on all spatial lattice sites, ii) methods that increase statistics at low cost e.g by using low precision inversions  (truncated solver method SM) or all-mode-averaging (AMA)), and iii) take advantage of graphics cards (GPUs) by developing special multi-GPU codes~\cite{Alexandrou:2012zz,Alexandrou:2013wca,Alexandrou:2014eva,Alexandrou:2014fva}.
 As an illustration of such a computation we consider one ensemble of $N_f=2+1+1$ 
 twisted mass fermions with lattice spacing $a = 0.082$~fm and $m_\pi$ = 373~Me and perform a high statistics analysis using $\sim 150, 000$ measurements for all  disconnected contributions to nucleon observables.
In Fig.~\ref{fig:gA disc} we show the ratio from which the disconnected contributions to $g_A^{u+d}$ and $g_A^s$ are extracted. These quantities  determine  the quark intrinsic spin $\Delta\Sigma^q$. As can be seen,  the disconnected  contributions are negative and non-zero and must be taken into account when
computing $\Delta \Sigma^q$. These results are in agreement with those by QCDSF~\cite{QCDSF:2011aa}. 
The spin in the nucleon satisfies the sum rule
 $ \frac{1}{2}=\sum_{q}\left(\frac{1}{2}\Delta \Sigma^q +L^q\right) +J^G $, where the quark contributions $J^q=\frac{1}{2}\Delta \Sigma^q +L^q$ can be computed from 
the relation $J^q=\frac{1}{2}\left(A_{20}^q(0)+B_{20}^q(0)\right)$. Furthermore knowing   $\Delta \Sigma^q=g_A^q$ we can extract the angular momentum $L^q$.
In Fig.~\ref{fig:quark spin} we show results on $J^{u,d}$, $\Delta \Sigma^{u+d}$ and $L^{u+d}=J^{u+d}-\Delta \Sigma^{u+d}$ neglecting disconnected contributions except at $m_\pi=373$~MeV where we also include the result after adding the disconnected contribution that leads to a decrease of $\Delta \Sigma^{u+d}$. 
What these results show is that the disconnected contributions amount to a
$\sim 10$\% correction at $m_\pi\sim$ 370~MeV and must be included if we aim at a few percent accuracy. Also we find that  $\sum_{q=u,d,s}J^q \sim 1/4$ so that the question regarding the other $\sim$50\% contribution to  the spin of the nucleon still remains open.

\subsection{ Nucleon  $\sigma$-terms}

The nucleon $\sigma$-term
$\sigma_{\pi N} \equiv \frac{(m_u+m_d)}{2}\langle N|{\bar u}u+\bar{d}d|N\rangle$ measures the explicit breaking of chiral symmetry and it is an important phenomenological quantity 
extracted from analysis of low-energy pion-proton scattering data.
Its value, determining the  Higgs-nucleon coupling, represents the  largest uncertainty in interpreting experiments for dark matter searches~\cite{Ellis:2008hf}.
In lattice QCD it can be computed using the Feynman-Hellman theorem via $\sigma_{q}=m_q\frac{\partial m_N}{\partial m_q}$.
A measure for  the strange quark content of the nucleon is the ratio $y_N=\frac{2\langle N|\bar{s}s|N\rangle}{ \langle N|{\bar u}u+\bar{d}d|N\rangle}=1-\frac{\sigma_0}{\sigma_{\pi N}}$, where $\sigma_0=\langle N|{\bar u}u+\bar{d}d-2{\bar s}s|N\rangle$ is the flavor non-singlet.
A number of groups have used the spectral method to extract the $\sigma$-terms~(see e.g. \cite{Young:2013nn}). However, we can now also calculate them directly by computing the
three-point functions including the disconnected contributions. In Fig.~\ref{fig:sigma} we show $y_N$ as a function of $m_\pi$  and extrapolated to the physical pion mass. Due to the cancellation of lattice systematics in the ratio, $y_N$ can be computed to a better accuracy than the $\sigma$-terms. 
We find $y_N=0.135(22)(33)(22)(9)$ where the first error is statistical, the second error is estimated using lowest order and next to leading order chiral perturbation theory for the chiral extrapolation, the third is due to excited states contamination and the fourth is an estimate of cut-off effects.  Using our value of $y_N$ and the phenomenological constrains on $\sigma_{\pi N}$  we can put a bound on the value of $\sigma_s\stackrel{<}{\sim} 250$~MeV~\cite{Alexandrou:2013nda} .

\section{Conclusions}

    Nucleon structure is a benchmark for lattice QCD calculations and thus the
 investigation of  $g_A$, $\langle x \rangle _{u-d}$,  $\langle x \rangle _{\Delta u-\Delta d}$ is considered a central issue. Simulations  at the physical pion mass and larger volumes are now becoming available and thus we expect lattice QCD to resolve any remaining discrepancies  by using  high statistics analysis and careful cross-checks. 
The evaluation of disconnected quark loop diagrams has also become feasible thus addressing an up to now unknown systematic error.
Reproducing the nucleon benchmark quantities will open the way for providing reliable predictions for other hadron observables such  axial charges and form factors of hyperons and charmed baryons.
Furthermore, appropriate methods to  study of excited states, resonances and decays are being developed, with good prospect of
 providing insight into the structure of hadrons and input that is crucial for experimental searches for new physics.

\section{Acknowledgments}
I would like to thank my collaborators A. Abdel-Rehim, M. Constantinou, V. Drach, K. Hadjiyiannakou, K. Jansen, Ch. Kallidonis,  G. Koutsou, Th. Leontiou and A. Vaquero without whom this work would not be possible. 
I would also like to thank all members of ETMC for a very constructive and enjoyable collaboration and for the
many fruitful discussions.
Numerical calculations have used HPC resources from John von Neumann-Institute for Computing  at the research center in J\"ulich through the PRACE allocation,
 and from the Cy- Tera facility of the Cyprus Institute under the project Cy-Tera (NEA Y$\Pi$O$\Delta$OMH/$\Sigma$TPATH/0308/31). This work is supported in part by the Cyprus Research Promotion Foundation under contracts  TECHNOLOGY/$\Theta$E$\Pi$I$\Sigma$/0311(BE)/16 and $\Pi$PO$\Sigma$E$\Lambda$KY$\Sigma$H/EM$\Pi$EIPO$\Sigma$/0311/16, and the Research Executive Agency of the European Union under Grant Agreement number PITN-GA-2009-238353 (ITN STRONGnet).

\bibliography{refs}

\end{document}